\documentclass[apj,numberedappendix]{emulateapj}
\usepackage{apjfonts}

\shorttitle{G-MODES DURING SIMMERING} 
\shortauthors{PIRO}

\newcommand{\be}{\begin{eqnarray}}
\newcommand{\ee}{\end{eqnarray}}
\newcommand{\lp}{\left(}
\newcommand{\rp}{\right)}
\newcommand{\lb}{\left[}
\newcommand{\rb}{\right]}

\begin{document}


\slugcomment{Submitted for publication in The Astrophysical Journal Letters.}

\title{G-Mode Excitation During the Pre-explosive Simmering of Type Ia Supernovae}

\author{Anthony L. Piro}

\affil{Theoretical Astrophysics, California Institute of Technology, 1200 E California Blvd., M/C 350-17, Pasadena, CA 91125; piro@caltech.edu}

\begin{abstract}
Prior to the explosive burning of a white dwarf (WD) that makes a Type Ia supernova (SN Ia), the star ``simmers'' for $\sim 10^3$ yrs in a convecting, carbon burning region. I estimate the excitation of {\it g}-modes by convection during this phase and explore their possible affect on the WD. As these modes propagate from the core of the WD toward its surface, their amplitudes grow with decreasing density. Once the modes reach nonlinear amplitudes, they break and deposit their energy into a shell of mass $\sim10^{-4}M_\odot$. This raises the surface temperature by $\approx6\times10^8\ {\rm K}$, which is sufficient to ignite a layer of helium, as is expected to exist for some SN Ia scenarios. This predominantly synthesizes $^{28}$Si, $^{32}$S, $^{40}$Ca, and some $^{44}$Ti. These ashes are expanded out with the subsequent explosion up to velocities of \mbox{$\sim20,000\ {\rm km\ s^{-1}}$}, which may explain the high velocity features (HVFs) seen in many SNe Ia. The appearance of HVFs would therefore be a useful discriminant for determining between progenitors, since a flammable helium-rich layer will not be present for accretion from a C/O WD as in a merger scenario. I also discuss the implications of $^{44}$Ti production.
\end{abstract}



\section{Introduction}

The use of Type Ia supernovae (SNe Ia) as cosmological distance indicators has brought attention to the uncertainties that remain about these events. It is generally agreed that they result from the unstable thermonuclear ignition of a C/O white dwarf (WD), but the exact progenitor is still unclear. It may be a single degenerate \citep[WD accreting from an evolved companion;][]{wi73}, double degenerate \citep[merging of two WDs;][]{web84,it84,pac85}, or combination of both scenarios. Any observational or theoretical clues that could help unravel this mystery are extremely useful.

In cases when the WD first unstably ignites carbon at its center, it subsequently undergoes $\sim10^3\ {\rm yrs}$ of convective simmering before the explosive burning wave is born \citep{woo04,ww04}. In single degenerate scenarios, central ignition occurs when the accretion proceeds at rates slower than the WD thermal timescale \citep{her88}. Central ignition can also occur via focusing of a shock from a surface helium detonation, but this leads to core detonation and not convection \citep{fin10}. During the late stages of convection there is considerable luminosity going into convective motions ($\sim10^{45}\ {\rm ergs\ s^{-1}}$), which is much greater than the Eddington luminosity of a Chandrasekhar WD of $L_{\rm Edd}\approx2\times10^{38}\ {\rm ergs\ s^{-1}}$. This energy is expected to be bottled up within the convective region because the thermal conduction timescale is $\sim10^6\ {\rm yrs}$, which is much longer than the convective timescale. If just a small fraction of this energy could be transported closer to the WD surface, it might have an effect on the surface structure, and may even have important observable consequences.

In the present work I consider the stochastic excitation of {\it g}-modes by convection and how they may affect the WD surface. The presence of such modes was first suggested by \citet{pc08} and subsequently observed in the simulations of \citet{zin09}. In \S \ref{sec:gmodes}, I make analytic estimates for the luminosity and total integrated energy in {\it g}-modes. I show how the modes grow as they propagate toward shallower densities, and argue that the mode energy is deposited near the WD surface when the modes break due to reaching nonlinear amplitudes. In \S \ref{sec:surface}, I study the detailed structure of the heated WD surface layers. The temperature rises sufficiently to ignite a surface helium layer, and the resulting ashes are predominantly composed of $^{28}$Si, $^{32}$S, $^{40}$Ca, and perhaps $^{44}$Ti. This may explain the high velocity features \citep[HVFs,][]{maz05b} seen in many SNe Ia. In \S \ref{sec:conclusion}, I summarize this study and discuss where possible future work is needed.


\section{g-mode Excitation and Propagation}
\label{sec:gmodes}

Ignition of $^{12}$C occurs when the heating from carbon fusion overpowers neutrino cooling \citep{nom84}. The central temperature $T_c$ then rises and a convective core grows, eventually encompassing $\sim1M_\odot$ of the WD after $\sim10^3\ {\rm yrs}$. For a central density $\rho_c$, the energy generation rate from carbon burning is \citep{woo04}
\be
	\epsilon = 2.8\times10^{13}
		\lp\frac{T_{c,8}}{7}\rp^{23}\lp\frac{\rho_{c,9}}{2}\rp^{3.3}{\rm ergs\ g^{-1}\ s^{-1}},
\ee
where $T_{c,8}=T_c/10^8\ {\rm K}$ and $\rho_{c,9}\equiv\rho_c/10^9\ {\rm g\ cm^{-3}}$, and equal mass fractions of carbon and oxygen are assumed. The central temperature increases on a heating timescale $t_h= (d\ln T_c/dt)^{-1}$, which gets shorter as $T_c$ becomes larger, and generally depends on the size of the convecting region \citep{pc08}. The simmering ends once $t_h< t_c$, where $t_c$ is the eddy overturn timescale. At these late times, individual eddies may experience significant heating during their transit \citep{gw95}, and there is not sufficient time for the entire convective region to respond to the increasing $T_c$. In this case it can be approximated that $t_h\approx c_pT_c/\epsilon$, where $c_p$ is the specific heat capacity at constant pressure. The heat capacity in the WD core, including Coulomb corrections, is \mbox{$c_p \approx1.3\times10^7\ {\rm ergs\ g^{-1}\ K^{-1}}$}, which is used to find
\be
	t_h \approx3\times10^2\ {\rm s}\ 
		\lp\frac{T_{c,8}}{7} \rp^{-22}\lp\frac{\rho_{c,9}}{2} \rp^{-3.3}.
	\label{eq:theat}
\ee
Woosley et al. (2004), using the KEPLER code \citep{wea78}, estimate that convection ends when $T_c\approx7.8\times10^8\ {\rm K}$ and $\rho_c\approx2.6\times10^9\ {\rm g\ cm^{-3}}$. This gives $t_h\sim10\ {\rm s}$, roughly the convective overturn timescale as shown below.

Integrating over the burning region of the core, the total luminosity carried by convection is  \citep{woo04}
\be
	L_c \approx 7\times10^{44}\lp\frac{T_{c,8}}{7}\rp^{23}
		\lp\frac{\rho_{c,9}}{2}\rp^{4.3} {\rm ergs\ s^{-1}}.
\ee
The convective velocity at the largest scales is $V_c\approx (L/4\pi r^2\rho)^{1/3}$. What is crucial for driving {\it g}-modes is the properties near the top of the convective zone, which has density and radius $\rho_t$ and $r_t$, respectively. The velocity here is
\be
	V_c\approx 4\times10^6\rho_{t,8}^{-1/3}r_{t,8}^{-2/3}
	\lp\frac{T_{c,8}}{7}\rp^{7.7}\lp\frac{\rho_{c,9}}{2}\rp^{1.4} {\rm cm\ s^{-1}}.
\ee
where $\rho_{t,8}=\rho_t/10^8\ {\rm g\ cm^{-3}}$ and $r_{t,8}=r_t/10^8\ {\rm cm}$. The spectrum of {\it g}-modes excited by the convection is peaked at a frequency equal to the eddy turnover frequency $\omega_c\approx V_c/H_t$, where $H_t$ is the scaleheight at the top of the convection. Taking $H_t\approx2\times10^7\rho_{t,8}^{1/3}g_{10}^{-1}\ {\rm cm}$, where \mbox{$g_{10}=g/10^{10}\ {\rm cm\ s^{-2}}$}, results in
\be
	\omega_c \approx 0.1g_{10}\rho_{t,8}^{-2/3}r_{t,8}^{-2/3}
	\lp\frac{T_{c,8}}{7}\rp^{7.7}\lp\frac{\rho_{c,9}}{2}\rp^{1.4}\ {\rm s^{-1}}.
\ee
Note the convective timescale is $t_c\sim H_t/V_c\sim 10\ {\rm s}$, roughly in agreement with when the convection should end, as discussed above. These waves propagate in the non-convective WD surface layers if their frequency satisfies $\omega_c<N$, where $N$ is the Brunt-V\"{a}is\"{a}l\"{a} frequency. This is approximated as
\be
	N\approx\lp\frac{g}{H}\frac{k_{\rm B}T_t}{ZE_{\rm F}}\rp^{1/2}
	\approx 0.4g_{10} T_{t,8}^{1/2}\rho_8^{-1/3}{\rm s^{-1}},
\ee
where $\rho$ (no subscript) is the density at some position near the WD surface, $\rho_8=\rho/10^8\ {\rm g\ cm^{-3}}$, $k_{\rm B}$ is Boltzmann's constant, $T_t$ is the temperature at the top of the convection, $Z$ is the average charge per ion, and
$E_{\rm F}$ is the Fermi energy for a degenerate, relativistic electron gas.  I take $Z=13.8$, as is appropriate for a mixture of equal parts carbon and oxygen, and ignore the scalings with composition to simplify the presentation. Since $\omega_c<N$ at the convective boundary, and at shallower depths $N\propto\rho^{-1/3}$, the {\it g}-modes propagate freely toward the surface.

The fraction of $L_c$ that can be put into {\it g}-modes is directly proportional to the Mach number of the convective eddies near the top of the convective zone \citep{gk90}\footnote{Some energy is expected to go into {\it p}-modes as well, but because the efficiency is $\propto Ma^{15/2}$, this is negligible.}. For a soundspeed $c_s=(4P/3\rho)^{1/2}$,
\be
	Ma = \frac{V_c}{c_s} \approx 7\times10^{-3}\ \rho_{t,8}^{-1/2}r_{t,8}^{-2/3}
	\lp\frac{T_{c,8}}{7}\rp^{7.7}\lp\frac{\rho_{c,9}}{2}\rp^{1.4}.
\ee
The {\it g}-mode luminosity is then
\be
	L_g &\approx& Ma L_c
	\nonumber
	\\
	&\approx& 5\times10^{42}\rho_{t,8}^{-1/2}r_{t,8}^{-2/3}
	\lp\frac{T_{c,8}}{7}\rp^{30.7}\lp\frac{\rho_{c,9}}{2}\rp^{5.7}
	{\rm ergs\ s^{-1}}.
	\nonumber
	\\
	\label{eq:lg}
\ee
Comparing the dependence on $T_c$ in equations (\ref{eq:lg}) and (\ref{eq:theat}) shows that $L_g\propto t_h^{-1.4}$. The total amount of energy put into {\it g}-modes up to any given time is therefore
\be
	E_g &=& \int L_g dt\approx2.5 L_g t_h
	\nonumber
	\\
	&\approx& 4\times10^{45}
	\rho_{t,8}^{-1/2}r_{t,8}^{-2/3}
	\lp\frac{T_{c,8}}{7}\rp^{8.7}\lp\frac{\rho_{c,9}}{2}\rp^{2.4}
	{\rm ergs}.
	\label{eq:eg}
\ee
Depending the final $T_c$, about \mbox{$\sim10^{46}\ {\rm ergs}$} goes into {\it g}-modes.

As the {\it g}-modes propagate into the non-convective surface layers, they satisfy the dispersion relation
\be
	\omega_c^2=\frac{k_h^2}{k_r^2+k_h^2}N^2,
\ee
and have a group velocity of $V_g=\omega_c/k_r$, where
$k_r$ and  $k_h$ are the radial and horizontal wavenumbers, respectively. This relation does not include rotational modifications. The radial wavenumber is unaffected by the Coriolis force if the spin $\Omega$ is small in comparison to the buoyancy \citep{cl70,bg94}
\be
	\Omega \lesssim N^2 H/\omega r \sim 0.3\ {\rm s^{-1}}.
	\label{eq:omega}
\ee
In this limit the mode equations can be simplified using the ``traditional approximation'' to separate vertical and horizontal parts. Although $k_r$ remains the same, $k_h$ can depend on the angle with respect to the rotation axis, since the Coriolis force pushes the modes to be more concentrated near the equator \citep[see the angular eigenfunctions plotted in][]{pb04}. The spin of accreting WDs may indeed exceed that given by equation (\ref{eq:omega}), but  rotational modifications are ignored to simplify the current study.

The total luminosity in {\it g}-modes, $L_g$, is carried by $n$ modes, each with a characteristic Lagrangian displacement $\xi$. Assuming that the modes are excited with random phase, the luminosity is related to the size of the perturbations via
\be
	L_g \approx 4\pi r^2 \rho (\omega \xi)^2 n V_g.
	\label{eq:lgmodes}
\ee
From incompressibility, the components of $\xi^2=\xi^2_r+\xi^2_h$ are related by $(\xi_r/\xi_h)^2\approx(k_h/k_r)^2$. The linearity of the modes is best represented by the dimensionless quantity $k_r\xi_r$, since both density inversion instabilities and Kelvin-Helmholtz instability set in when $k_r\xi_r\sim O(1)$ \citep{mt81}. Using the above relations,
\be
	k_r\xi_r=\lp \frac{L_g}{4\pi r^2\rho n}\rp^{1/2}
	\frac{k_h^{3/2}}{N\omega^{1/2}}\lb\lp\frac{N}{\omega}\rp^2-1 \rb^{3/4}.
\ee
The number of modes is found by integrating over the mode spectrum and the surface area of the convection. There remain uncertainties about the power spectrum of the convection \citep[see the discussions in][]{woo04,kwg06}, which in turn determines the spectrum of modes. It is therefore sufficient to estimate
\be
	n\sim (r k_h)^2,
\ee
keeping in mind that this is a lower limit since a wider spectrum will require more modes to carry the same luminosity. In the limit $\omega\ll N$, the amplitude simplifies to
\be
	k_r\xi_r \approx \lp \frac{L_g}{4\pi r^2\rho}\rp^{1/2}
	\frac{k_h^{1/2}N^{1/2}}{r\omega^2},
\ee
and taking $k_h\approx 1/H_t$, I evaluate this as a function of $\rho$,
\be
	k_r\xi_r \approx 3\times10^{-2} g_{10}^{-1}\rho_{t,8}^{11/12}r_{t,8}^{-1}T_{t,8}^{1/4}\rho_8^{-2/3}.
\ee
As the mode propagates toward the surface $k_r\xi_r\propto \rho^{-1/2}N^{1/2}\propto \rho^{-2/3}$, so this dimensionless amplitude grows. This quantity is independent of the strength of convection, as apparent from the lack of a dependence on $T_c$ or $\rho_c$. This is because the amplitude is \mbox{$k_r\xi_r\propto L_g^{1/2}\omega^{-2}$}. As the driving becomes more vigorous, the {\it g}-mode luminosity increases, but so does the frequency, and the two effects exactly balance\footnote{In detail, as the convection proceeds, $r_t$ will increase, while $g$ and $\rho_t$ will decrease, but these are small correction with respect to the simplifying assumptions I am making here.}.

Near the surface the thermal conduction timescale becomes shorter. The thermal conductivity from electron-ion scattering is (Yakovlev \& Urpin 1980)
\be
	K = \frac{\pi^2 k_{\rm B}^2Tn_e}{3m_*\nu},
\ee
where $m_*\approx E_{\rm F}/c^2$, $n_e$ is the number density of electrons, and $\nu = 4m_*Ze^4 \Lambda/4\pi \hbar^3$ is the electron-ion collision frequency with $\Lambda\approx1$ is a weakly dependent function of density in the ocean and $\hbar$ is Planck's constant. For a lengthscale $\lambda$, the local thermal time is $t_{\rm th}\sim \rho c_p \lambda^2/K$, giving
\be
	t_{\rm th}\approx 7\times10^4\rho_8^{2/3} \lp \frac{\lambda}{2\times10^7\ {\rm cm}}\rp^2 {\rm yrs},
	\label{eq:tth}
\ee
where $\lambda$ has been scaled to the typical horizontal wavelength of the {\it g}-modes. This is much longer than the timescale for a mode to travel a scaleheight $t_g \approx H/V_g \approx (H/H_t)N/\omega_c\sim 40\ {\rm s}$, so conduction does not damp the modes. Since breaking occurs before damping, setting $k_r\xi_r\approx1$, I find
\be
	\rho_b \approx 5\times10^5g_{10}^{-3/2}\rho_{t,8}^{11/8}r_{t,8}^{-3/2}T_{t,8}^{3/8}\ {\rm g\ cm^{-3}}.
	\label{eq:rhob}
\ee
for the mode breaking density.

\section{Surface Heating}
\label{sec:surface}

For the breaking depth found in equation (\ref{eq:rhob}), the corresponding mass of material is $M_b\sim 4\pi r^2 \rho_bH(\rho_b)\sim10^{-4}\ M_\odot$. If an energy $E_g$ is put into a shell with mass $M_b$, it does not eject the material, since the binding energy for a Chandrasekhar WD is much greater at $GMM_b/R\sim 10^{47}\ {\rm ergs}$. Instead the energy input heats the surface, which I now explore.

A simple estimate for the change of temperature is $\Delta T \sim E_g/c_p M_b\sim10^9\ {\rm K}$, but this does not take into account how the heating is distributed. The heating at a depth $\rho_b$ acts like a hot plate, which still cannot be carried by conduction due to the long thermal timescale (see eq. [\ref{eq:tth}], evaluated at $\rho_b$). Instead, a secondary convective zone grows, and since the heating is now distributed over this entire new convective region, the temperature rise at the base will in general be less than the $\Delta T\sim10^9\ {\rm K}$ estimated above.

\begin{figure}
\epsscale{1.2}
\plotone{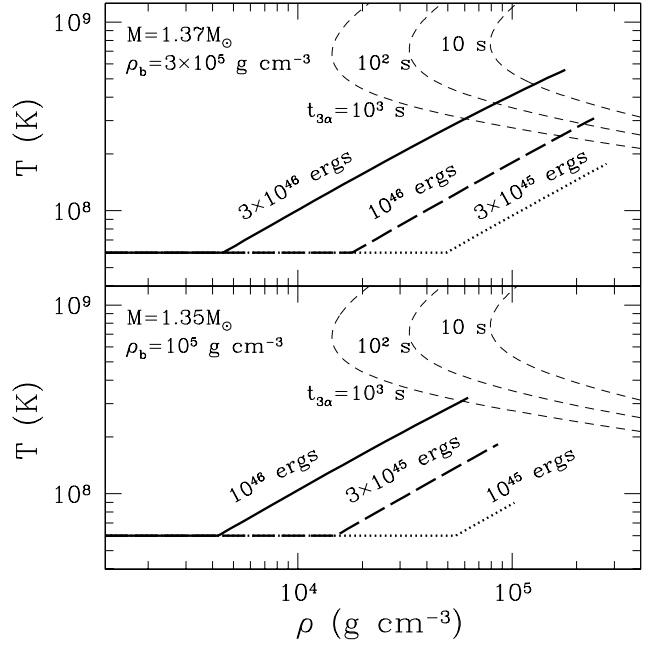}
\caption{Example temperature profiles for the surface convection zone caused by
energy deposited from {\it g}-modes. In the top panel I use $M=1.37M_\odot$, $R=1.6\times10^8\ {\rm cm}$, and $\rho_b=3\times10^5\ {\rm g\ cm^{-3}}$, and in the bottom panel $M=1.35M_\odot$, $R=2.1\times10^8\ {\rm cm}$, and $\rho_b=10^5\ {\rm g\ cm^{-3}}$. The thick lines show convective profiles for different amounts of total internal energy, as labeled. The thin, short-dashed lines are curves of constant $t_{3\alpha}$, demonstrating when helium burning begins.}
\label{fig:surface}
\epsscale{1.0}
\end{figure}

To better understand the extent and peak temperature of this secondary convection zone, I build a series hydrostatic surface models where the convection zone is treated as an adiabat. The initial model is assumed isothermal with a temperature of $T_i=6\times10^7\ {\rm K}$, similar to the surface temperature in accreting WD models \citep{yl04}. The convection extends from a bottom mass coordinate $M_1$ set by the {\it g}-mode breaking depth up to top mass coordinate $M_2$ set by the total energy injected by the {\it g}-modes
\be
	E_g = \int_{M_1}^{M_2} c_p [T(M_r)-T_i] dM_r,
	\label{eq:energyintegral}
\ee
where $T(M_r)$ is the temperature profile. This integral is fairly insensitive to the assumption of $T_i$, since at late times the convection has much more energy than the initial temperature profile. Even though $M_1$ is set fixed by the mode breaking depth, the density at the base of the convective zone decreases because of the increasing temperature at fixed pressure.

In Figure \ref{fig:surface}, I plot example thermal profiles that would be present from energy deposited by {\it g}-modes. In the top panel I use $M=1.37M_\odot$, $R=1.6\times10^8\ {\rm cm}$, and $\rho_b=3\times10^5\ {\rm g\ cm^{-3}}$, and in the bottom panel $M=1.35M_\odot$, $R=2.1\times10^8\ {\rm cm}$, and $\rho_b=10^5\ {\rm g\ cm^{-3}}$. The central densities of the two models are $6.1\times10^9\ {\rm g\ cm^{-3}}$ and $2.0\times10^9\ {\rm g\ cm^{-3}}$, respectively. The difference in breaking depths was estimated from core-convective models \citep[as calculated in][]{pb08,pir08}. Each thick line shows the thermal profile for a labeled amount of {\it g}-mode energy $E_g$, solved using equation (\ref{eq:energyintegral}). The more massive WD is plotted with a larger $E_g$ due to its higher central density (eq. [\ref{eq:eg}]). The base temperature rises by $\Delta T\approx 6\times10^8\ {\rm K}$ and for the top panel and $\approx 3\times10^8\ {\rm K}$ for the bottom panel, but for a given $E_g$, $\Delta T$ is roughly the same. These detailed profiles also provide an estimate of the position of the top of the convection zone, which is near $\sim\ {\rm few}\times10^3\ {\rm g\ cm^{-3}}$. If the thermal time here is sufficiently short, heat leaves the top of the convection and stunts its growth, as is found for Type I X-ray bursts \citep{wb06}. Since the thermal time $t_{\rm th}\sim 1\ {\rm yr}$ (eq. [\ref{eq:tth}]) is much longer than $t_h$,  conduction does not truncate the secondary convective zone.

Besides changing the entropy and thermal profile, the heating may have an additional effect if the surface layers are rich in helium. A thin layer of helium is expected in many single degenerate scenarios that require accretion to grow to a Chandrasekhar mass. For an energy generation rate from triple-$\alpha$ reactions of $\epsilon_{3\alpha}$, the timescale for increasing the temperature is \mbox{$t_{3\alpha}\approx c_p T/\epsilon_{3\alpha}$}. Figure \ref{fig:surface} plots $t_{3\alpha}=10, 10^2,$ and $10^3\ {\rm s}$ ({\it thin, short-dashed lines}), using the reaction rates from \citet{fl87}. Comparison of these curves with the convective profiles demonstrates that triple-$\alpha$ reactions can begin burning the layer. Helium burning favors deeper mode breaking and more massive WDs, since a higher density leads to a shorter $t_{3\alpha}$. A more detailed study is required to assess the full range of WD masses and breaking depths needed for helium ignition, which is outside the scope of this work.

The low shell mass is below what is needed for dynamical burning \citep{sb09}, so a detonation is not expected. Nevertheless, energy from helium-burning provides \mbox{$\approx5.8\times10^{17}\ {\rm ergs\ g^{-1}}$}, similar to the binding energy of a Chandrasekhar WD ($\sim10^{18}\ {\rm ergs\ g^{-1}}$), so it is possible that some material is expelled.  A breaking depth of $\rho_b\approx(3-5)\times10^5\ {\rm g\ cm^{-3}}$ corresponds to a base pressure of $\approx(4-9)\times10^{21}\ {\rm ergs\ cm^{-3}}$. The burning of helium at constant pressure was explored by \citet{has83}, the products of which are summarized in their Figure 10. For this pressure range, the predominant elements are $^{28}$Si, $^{32}$S, and $^{40}$Ca, but up to a mass fraction of $\sim0.1$ of  $^{36}$Ar and $^{44}$Ti may also be present depending on $\rho_b$.

The composition and position of these elements are similar to what is required for the HVFs seen in many (or perhaps most) SNe Ia \citep{maz05b}.  HVFs most prominently show $^{40}$Ca, although $^{28}$Si is sometimes present. Currently explanations for these features include circumstellar interactions \citep{ger04}, three-dimensional density or composition enhancements \citep{max05a}, or ashes on the surface from a gravitationally confined explosion \citep{kp05}. Modeling of the early time lightcurve, including the HVFs was done by \citet{tan08}. Surface burning from {\it g}-mode energy input suggests that $^{32}$S is expected to be fairly abundant as well, but the physics of its line creation may not be optimal for seeing it. Nevertheless, searching for signs of $^{32}$S would be an important test of this hypothesis. If this model is correct, the presence of HVFs would be an indication that helium accretion (or hydrogen accretion which then burns to form helium) is responsible for these SNe Ia, which is expected for a single degenerate progenitor or a degenerate helium donor (like for AM CVn stars).


\section{Conclusion and Discussion}
\label{sec:conclusion}

I considered the driving of {\it g}-modes via convection during the simmering stage prior to SNe Ia. I found that the {\it g}-modes are driven with a luminosity $L_g\gtrsim10^{42}\ {\rm ergs\ s^{-1}}$, injecting $E_g\sim10^{46}\ {\rm ergs}$ at a density of $\rho_b\sim\ {\rm few}\times10^5\ {\rm g\ cm^{-3}}$ where the modes break. I considered the effect of this energy in generating a secondary, surface convective region and showed it reaches base temperatures of $\approx6\times10^8\ {\rm K}$, sufficient to ignite helium. The ashes expected from this burning are similar to the elements observed in HVFs, which may mean that HVFs indicate a progenitor channel different from the merger of two C/O WDs since these would not have helium present.

This study motivates further, more detailed investigations. The mode excitation estimates employed here make use of studies done in the context of the Sun \citep{gk90}, and may need to be modified for a degenerate equation of state. In addition, rotation can strongly influence the morphology of WD convection \citep{kwg06}, which would in turn alter the mode wavelengths and spectrum (also see the discussion of rotational modifications to the modes in \S 2). Rotation also breaks the spherical symmetry of the WD and would imprint asymmetries into the flux of {\it g}-modes.

A study of thin helium-shell burning on massive WDs would provide detailed predictions for the burning products and determine what fraction of material can be expelled. The production of $^{44}$Ti is especially interesting for testing this model and investigating SNe Ia progenitors. Our fiducial calculations estimate $\sim10^{-5}M_\odot$ of $^{44}$Ti is produced, although this varies strongly as a function of $\rho_b$. \citet{bork10} find $(1-7)\times10^{-5}M_\odot$ of $^{44}$Ti in the remnant G1.9+0.3 from the decay product $^{44}$Sc. The $^{44}$Sc is most abundant in the remnant's north rim, which is surprising if the $^{44}$Ti is expected to be synthesized in the neighborhood of more centrally located Fe-peak elements \citep{iwa99}. Further studies of the distribution of $^{44}$Sc would help determine whether surface helium burning is a viable alternative. Detections of the $68\ {\rm keV}$ decay line from $^{44}$Ti in SNe Ia by future satellites like NuSTAR \citep{har10} may also provide an important constraint.

\acknowledgments
I thank Philip Chang for some of the original conversations that inspired this research, and I thank Lars Bildsten, Peter Goldreich, Fiona Harrison, Paolo Mazzali, Christian Ott, Eliot Quataert, Nevin Weinberg, and Stan Woosley for helpful feedback. I also acknowledge the support of the Astronomy Department at UC Berkeley, where some of this research was conducted. This work was supported through NASA ATP grant NNX07AH06G, NSF grant AST-0855535, and by the Sherman Fairchild Foundation.


\end{document}